\documentclass[english,a4paper]{article}
\usepackage{dcolumn}
\usepackage{bm}
\usepackage{graphicx}
\usepackage{amsmath}
\usepackage{amsfonts}
\usepackage{amssymb}
\usepackage{amsthm}
\usepackage{amstext}
\usepackage{amsbsy}
\usepackage{amsopn}
\usepackage{amscd}
\usepackage{amsxtra}

\def\openone{\leavevmode\hbox{\small1\kern-3.3pt\normalsize1}}

\usepackage[usenames,dvipsnames]{color}

\newcommand{\bec}{\begin{center}}
\newcommand{\enc}{\end{center}}
\newcommand{\be}{\begin{equation}}
\newcommand{\ee}{\end{equation}}
\newcommand{\bmi}{\begin{minipage}}
\newcommand{\emi}{\end{minipage}}

\newcommand{\bi}{\begin{itemize}}
\newcommand{\ei}{\end{itemize}}
\newcommand{\ba}{\begin{array}}
\newcommand{\ea}{\end{array}}

\newcommand{\lgr}{\left\{}

\newcommand{\rd}{\right.}

\begin{document}

\title{Optimal control theory for applications in Magnetic Resonance Imaging}
\author{E. Van-Reeth, H. Ratiney\footnote{CREATIS, Universit\'e Lyon 1, 3 rue Victor Grignard, Villeurbanne, France}, M. Lapert, S. J. Glaser\footnote{Department of Chemistry, Technische Universit\"at
M\"unchen, Lichtenbergstrasse 4, D-85747 Garching, Germany} and D. Sugny\footnote{Laboratoire Interdisciplinaire Carnot de
Bourgogne (ICB), UMR 6303 CNRS-Universit\'e Bourgogne-Franche Comt\'e, 9 Av. A.
Savary, BP 47 870, F-21078 Dijon Cedex, France and Institute for Advanced Study, Technische Universit\"at M\"unchen, Lichtenbergstrasse 2 a, D-85748 Garching, Germany, dominique.sugny@u-bourgogne.fr}}

\maketitle

\begin{abstract} % abstract
We apply innovative mathematical tools coming from optimal control
theory to improve theoretical and experimental techniques in Magnetic Resonance Imaging (MRI).
This approach allows us to explore and to experimentally reach the physical limits of the corresponding spin dynamics in the presence of
typical experimental imperfections and limitations. We study in this paper two important goals, namely the optimization
of image contrast and the maximization of the signal to noise per unit time. We anticipate that the proposed
techniques will find practical applications in medical imaging in a near future to help the
medical diagnosis.
\end{abstract}

\section{Introduction}
Optimality with respect to a given criterion is vital in many applications, but it presents a complexity that requires a lot of ingenuity to provide a solution. In this context, optimal control tackles the question of bringing a dynamical system from one state to another with minimum expenditure of time and resources \cite{Pontryagin1962}. Optimal control theory was born in its modern version with the Pontryagin Maximum Principle (PMP) in the late 1950's. Its
development was originally inspired by problems of space dynamics, but it is now a key tool to
study a large spectrum of applications extending from robotics to economics and biology \cite{Bonnard2003,boscainbook,DAlessandro2008,Jurdjevic1997,sugnybook,Brif2010,cat,Dong2010,DAlessandro2001}. Optimal control problems can be solved by two different types of approaches, geometric \cite{Bonnard2003,boscainbook,Jurdjevic1997}
and numerical methods \cite{cat,brysonbook,Krotov1996} for dynamical systems of low and high dimension,
respectively. The geometric techniques lead to a complete mathematical and qualitative understanding of the control problem, from which we can deduce the structure of the optimal solution and the physical limits of a dynamical
process, such as the minimum time to achieve a given task~\cite{Bonnard2003,boscainbook,Jurdjevic1997,Assemat2010,Garon2013,Kontz2008,Khaneja2001,Boscain2006,Khaneja2002}. The
advantages of the numerical approach are complementary to the ones of the geometric method.
The relative simplicity of the application of the numerical algorithms makes it possible to adapt
them straightforwardly to new classes of control problems. Generally, it is also possible to include
constraints in the algorithms to account for experimental imperfections or requirements related to a
specific material or device. This point is essential in view of experimental applications and helps to bridge the gap between
control theory and control experiments~\cite{cat,brysonbook,Krotov1996,Reich2012,Khaneja2005,Machnes2011,Fouquiere2011,Tosner2009}.

Since its discovery in the forties, Magnetic Resonance has become a
powerful physical tool to study molecules and matter in a variety of domains in chemistry, biology and solid state physics \cite{Ernst1987,Levitt2008,Bernstein2004}. The efficiency of Magnetic Resonance techniques is maybe best illustrated by
medical imaging, where it is now possible to build up a three-dimensional picture of the human
brain. Even the mental processes of the brain activity, which modify the oxygenation and the flow
of the blood can be detected by this tool. This information is not
accessible by any other current method \cite{Bernstein2004}. It is this imaging aspect of the control of spin dynamics
which will be at the core of this paper. In Magnetic Resonance Imaging (MRI), the major
challenges are based on the maximization of the contrast between volume elements with different
physical or chemical properties in order to clearly detect anatomical features and to distinguish
healthy tissue e.g. from tumors. It is also crucial to improve sensitivity by compensating
instrumental imperfections such as the non-uniformity of signal excitation magnetic pulses. Such goals can be achieved by applying specific magnetic fields to the sample.
The development of theoretical techniques, such as composite and shaped pulses in MRI has so far relied on a relatively small number of concepts and numerical
tools that have led to many practical improvements in the last decades \cite{Levitt2008}. However, these
improvements were largely incremental, none of those approaches was able to establish physical
limits of the best possible performance in terms of energy deposition, sensitivity, robustness and
contrast. The revolutionary aspect of optimal control theory in this domain is in its ability to achieve
and operate at the physical performance limit, resulting in more detailed information, better
sensitivity, and better imaging contrast. Measurement time per patient could also be
decreased, making the technique more patient-friendly and also more economical. A crucial issue
for pulse design is the implementation of control operations under significant uncertainties
stemming from the non-uniform nature of the sample, such as the human body, and of the applied
magnetic fields. Recent advances in numerical optimization techniques have made it possible to
design high quality control sequences that are robust against the various experimental
uncertainties~\cite{cat,Khaneja2005,Nielsen2010,Daems2013,Ruschhaupt2012}. This requirement of robustness in general is one of the key factors for the
development of new technologies. Recently it has been shown that
thousands of individual spins with different experimental features can be simultaneously optimized,
resulting in an improved experimental performance \cite{Skinner2003,Skinner2004,Skinner2005,Skinner2006,Kobzar2004,Kobzar2008,Kobzar2012,Gershenzon2007,Zhang2011,Lapert2012,Lapert2012b}. More
generally, the flexibility of the optimal control approach allows us to include complex physical and
instrumental effects, such as relaxation, radiation damping and power limits in order to
find highly robust experimental settings suitable for practical applications under realistic conditions.

The idea of using optimal control techniques in MRI was initially proposed in the eighties \cite{Conolly1986}, but the recent advent of analytical and numerical tools has made possible impressive progresses in this domain, leading to the optimization of the control of complex spin dynamics in MRI \cite{Lapert2012,Bonnard2012,Vinding2012,Aigner2016,Xu2008,Massire2013,Sbrizzi2016}. The aim of this paper is to present two examples of recent results obtained in this direction both from the geometric and numerical approaches. These examples will allow us to describe and discuss the efficiency and the limitations of this method in order to solve applied and concrete issues in MRI.

The paper is organized as follows. Section \ref{sec2} deals with the optimization of the signal to noise per unit time (SNR) by geometric control techniques. The unbounded case is investigated, showing that the Ernst angle solution is the optimal solution of the control problem. This result is an essential prerequisite for the extension of this analysis to realistic situations, which are numerically much more intricate. Section \ref{sec3} focuses on the optimization of the contrast in \emph{in vivo} MRI. We numerically maximize the contrast of the brain of a rat. The experimental results show the efficiency of the applied magnetic fields. Conclusion and prospective views are given in Sec. \ref{sec4}.
\section{Optimal control of the signal to noise ratio per unit time}\label{sec2}
This paragraph deals with an application of geometric optimal control theory to MRI. Recently, this method has been successfully applied to a question of fundamental and practical interest in MRI, the maximization of the achievable signal-to-noise ratio per unit time (SNR) of a spin 1/2 particle \cite{Ernst1987,Levitt2008,Bernstein2004}. The SNR is practically enhanced in spin systems by using a multitude of identical cycles. In this periodic regime, the SNR increases as the square root of the number of scans. Each elementary block is composed of a detection time and of a control period where the spin is subjected to a radio-frequency magnetic pulse, this latter being mandatory to guarantee the periodic character of the overall process. A first solution to this problem was established in the sixties by R. Ernst and his co-workers \cite{ernst1966}. In this proposition, the control law is made of a $\delta$- pulse, featured by a specific rotation angle, and known as the \emph{Ernst angle solution}. This pulse sequence is routinely used in magnetic resonance spectroscopic and medical applications. Note also that optimal control theory was not used in Ref.~\cite{ernst1966}. We have revisited recently this question by applying the powerful tools of geometric control theory. In~\cite{Lapert2014}, we have shown in the case of unbounded controls that the Ernst angle solution is the optimal solution of this control problem. This analysis was generalized in \cite{Lapert2015} to spin dynamics in the presence of radiation damping effects and crusher gradients. We present in this section a brief review of these results.
\subsection{The model system}
We describe in this paragraph the model system used in the theoretical analysis. We consider
an ensemble of spin 1/2 particles subjected to a radio-frequency magnetic field, which is assumed to be homogeneous across the sample. The system is described by a magnetization vector $\vec{M}$ of coordinates $(M_x,M_y,M_z)$, whose dynamics is governed, in a given rotating frame, by the standard Bloch equations \cite{Ernst1987,Levitt2008,Bernstein2004}:
\be
\lgr\ba{rl}
\dot{M}_x &= -2\pi M_x/T_2 + \omega_y M_z\\
\dot{M}_y &= - 2\pi M_y/T_2 - \omega_x M_z\\
\dot{M}_z &= 2\pi (M_0 - M_z)/T_1 - \omega_y M_x + \omega_x M_y
\ea\rd
\ee
where $T_1$ and $T_2$ are respectively the
longitudinal and transverse relaxation constants. $M_0$ is the thermal
equilibrium of the magnetization and $(\omega_x,\omega_y)$ are the two control
amplitudes of the radio-frequency magnetic field. Normalizing the time with respect to the detection time $T_d$ (see below for a description of this parameter), $\tau=t/T_d$, and the amplitude of the magnetization vector with respect to $M_0$, $(x,y,z)=(M_x/M_0,M_y/M_0,M_z/M_0)$, we get:
\be
    \lgr\ba{rl}
        \dot{x} &= - \Gamma x + u_y z\\
        \dot{y} &= - \Gamma y - u_x z\\
        \dot{z} &= \gamma(1 - z) - u_y x + u_x y
    \ea\rd
\ee
where $\gamma =2\pi  T_d/T_1$, $\Gamma=2\pi  T_d/T_2$ and
$u_{x,y}=T_d \omega_{x,y}$. In the relevant physical case where $0\leq T_2\leq 2T_1$, it can be shown that $x^2+y^2+z^2\leq 1$ at any time, which defines the Bloch ball \cite{Lapert2013a}. In the new coordinates, the dynamics only depends on two parameters $\gamma$ and $\Gamma$ and the detection time is fixed to 1. The system admits a symmetry of revolution around the $z$- axis which allows us to consider by setting $u_y=0$ only a planar projection onto the $(y,z)$- plane \cite{Bonnard2009a,Bonnard2009}. The final equations used in this paper are then given by:
\be \label{eq3}
    \lgr\ba{rl}
        \dot{y} &=  - \Gamma y - u z\\
        \dot{z} &=  \gamma(1 - z) + u y
    \ea\rd
\ee
where $u$ stands for $u_x$.

\begin{figure}[htbp]
\centering\includegraphics[scale=0.7]{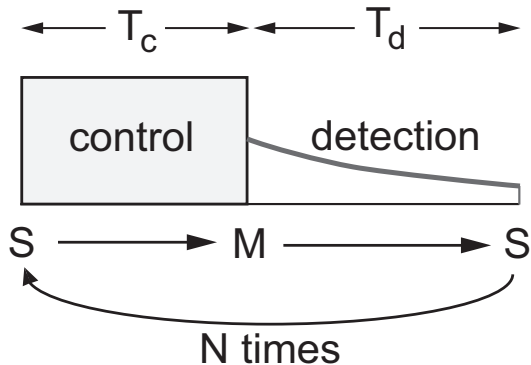}
\caption{\label{fig1} Schematic representation of the cyclic process used in the maximization of the SNR.}
\end{figure}

In this ideal dynamical system, we introduce a simple scenario to describe the optimization of the SNR per unit time (see the schematic description in Fig. \ref{fig1}. The point reached at the end of the pulse sequence is the measurement point $M$ of coordinates $(y_m,z_m)$. The system has a free evolution from this point to the steady state $S$ of coordinates $(y_s,z_s)$, where the pulse sequence starts. The times $T_d$ and $T_c$ denote the detection time (fixed by the experimental setup) and the control time, respectively. The total time $T$
during which a series of identical experiments are made is fixed. The total
number $N$ of experiments is then given by: $T=N(T_c+T_d)$.
Following Refs.~\cite{Lapert2014,Lapert2015}, the optimization problem is defined through the introduction of the following cost functional $\mathcal{R}$:
\be
  \mathcal{R}=\frac{Ny_m}{\sqrt{N}},
\ee
$y_m$ being the strength of the signal at the beginning of the detection
period (only the transverse magnetization can be measured). A white noise is assumed, which explains the $\sqrt{N}$ factor in the denominator of $\mathcal{R}$. Simple algebra leads to:
\be
\mathcal{R}=\sqrt{\frac{T}{T_d+T_c}}y_m=\sqrt{\frac{T}{T_d}}\frac{y_m}{\sqrt{1+T_c/T_d}}.
\ee
Since the total time $T$ and the detection time $T_d$ are fixed,we can introduce
a normalized SNR $Q$ with respect to the measurement time $T_d$ :
\begin{equation}\label{eqcost}
 Q(y_m,z_m)=\frac{y_m}{\sqrt{1+T_c(y_m,z_m)/T_d}},
\end{equation}
which only depends on the position of the $M$ point. The factor of quality $Q$
is the figure of merit to maximize.
\subsection{How to move in the Bloch ball}\label{move}
\begin{figure}[htbp]
        \centering\includegraphics[scale=0.5]{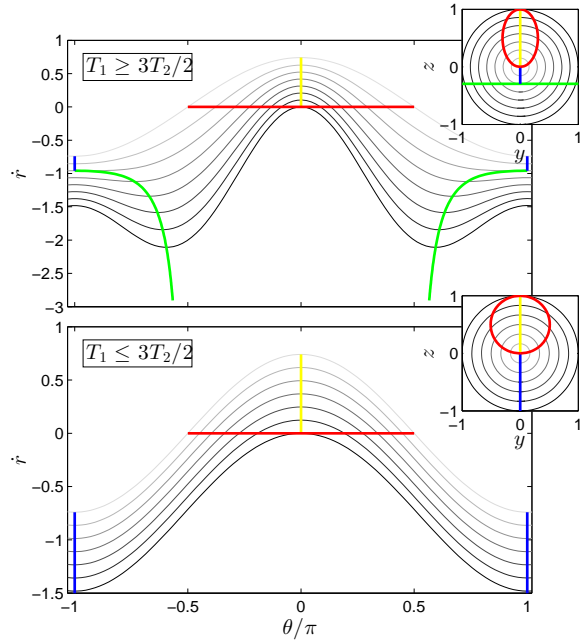}
        \caption{\label{fig2} (Color online) Contour plot of the trajectories $\dot{R}(\theta)$ for two different sets of relaxation parameters. The red line is the line for which $\dot{R}=0$. The blue and green lines are the set of points where $\dot{r}$ is minimum, while the points of the yellow one correspond to a maximum. The small inserts display in the $(y,z)$- plane  the trace of the corresponding sets of points. Circles of different radii are also plotted to help the reading.}
\end{figure}

\begin{figure}[htbp]
        \centering\includegraphics[scale=0.6]{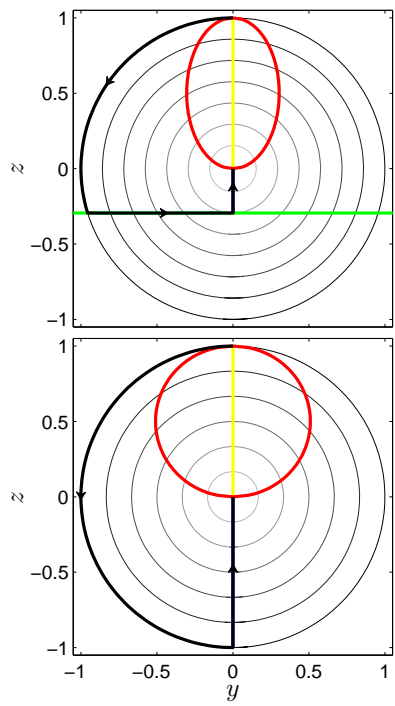}
        \caption{\label{fig3} (Color online) Time-optimal trajectory to reach the center of the Bloch ball $(y,z)=(0,0)$ from the north pole.}
\end{figure}

Before entering into the details of the optimization of the SNR, the first question to solve is the optimal control of the magnetization vector in the Bloch ball. Here, we only summarize the main results, which have been recently established in a series of papers \cite{Lapert2013a,Bonnard2009,Lapert2010,Lapert2011,Mukherjee2013}. We recall that the dynamics of the system is governed by Eq.~(\ref{eq3}), with no constraint on the control field, $u(t)$. At this stage, it is illuminating to introduce the polar coordinates $(R,\theta)$ such that $y=R\sin\theta$ and $z=R\cos\theta$. Equations~(\ref{eq3}) can then be written as follows:
\be \label{eq7}
    \lgr\ba{rl}
        \dot{R} &= -\Gamma R\cos^2\theta+\gamma (\sin\theta-R\sin^2\theta) \\
        \dot{\theta} &= \gamma\cos\theta (\frac{1}{R}-\sin\theta)+\Gamma\sin\theta\cos\theta+u.
    \ea\rd
\ee
We immediately note that the angular speed $\dot{\theta}$ can be directly control by $u$, while the radial one, $\dot{R}$, does not explicitly depend on the field. The radial coordinate can only be controlled in a two-step process by a judicious choice of $\theta$. In the unbounded case, we have therefore a complete control over the angular degree of freedom, but we still have to understand how to manipulate the system along the radial direction. For that purpose, we have plotted in Fig.~\ref{fig2} the evolution of the speed $\dot{R}$ as a function of the angle $\theta$. The characteristic points of this dynamics, maximum, minimum and zero points, are underlined and reported in the $(y,z)$- plane. Such specific points allow us to determine the optimal solution to manipulate in minimum time the system along the radial direction \cite{Lapert2013a}. From a control point of view, this analysis allows to recover the singular trajectories, which are the paths where the control field is not equal, in absolute value, to the maximum bound \cite{Lapert2010}. We obtain here two types of singular controls: the horizontal one along the plane of equation $z=z_0=\frac{-\gamma}{2(\Gamma-\gamma)}$, denoted $S_h$ and the vertical one along the $z$- axis denoted $S_{v>0}$ and $S_{v<0}$, for $z>z_0$ and $z<z_0$, respectively. Note that the plane of equation $z=z_0$ intersects the Bloch ball if
$$
\frac{T_2}{2|T_2-T_1|}\leq 1.
$$
In the standard case where $T_1\geq T_2$, this leads to the condition $2T_1\geq 3T_2$. The time-optimal solution is thus the concatenation of bang arcs, denoted $B$ where the intensity of the field is maximum (here infinite) and of singular arcs. We refer the reader to previous works for a thorough mathematical and physical analysis of this control process \cite{Lapert2013a,Bonnard2009,Lapert2010,Lapert2011}, but according to the constraint given below, the horizontal singular arcs can only be used if the condition $2T_1\geq 3T_2$ is satisfied. An example is given in Fig.~\ref{fig3} for the saturation process where the goal is to reach as fast as possible the center of the Bloch ball. According to the values of the relaxation parameters $T_1$ and $T_2$, we observe that two different control sequences are used. In the first situation, the optimal control law is composed of a bang pulse followed by two singular arcs, a horizontal and a vertical one. In the second case, an inversion pulse is used to reach the south pole before the application of a zero control along the $z$- axis. It can be shown that the analytical expression of the corresponding minimum times are given by:
$$
T_{min}=\frac{T_2}{2}\ln (1-\frac{2}{\alpha T_2})+T_1\ln (\frac{2T_1-T_2}{2(T_1-T_2)}),
$$
where $\alpha=\frac{T_2(T_2-2T_1)}{2T_1(T_1-T_2)^2}$
and
$$
T_{min}=T_1 \ln 2
$$
in the first and second cases, respectively.

A straightforward generalization of this study allows us to design the optimal field bringing in minimum time the system from any initial point to any target state of the Bloch ball, leading to a complete description and understanding of this optimal control problem. Such a classification of all the optimal trajectories is called an optimal
synthesis \cite{boscainbook}. In each case, we recall that the time-optimal solution can only be the concatenation of different bang arcs and of singular paths along the $z$- axis or along the line $z=z_0$.
\subsection{Optimization of the SNR in the unbounded case}
\begin{figure}[htbp]
        \centering\includegraphics[scale=0.5]{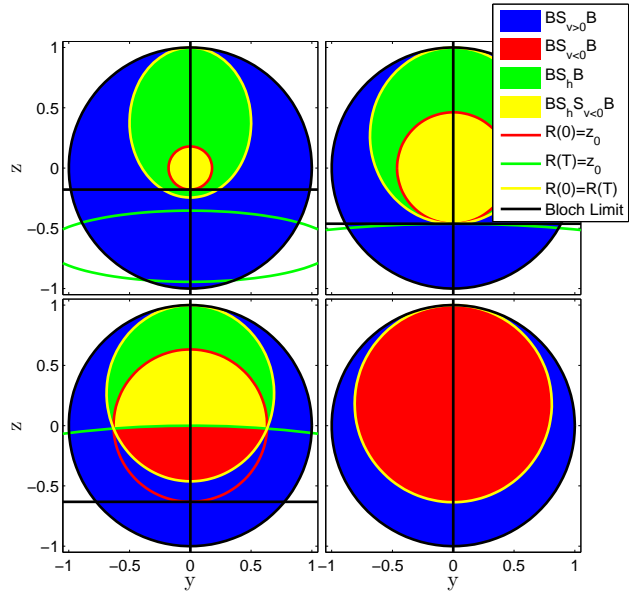}
        \caption{\label{fig4}(Color online) (Color online) Plot in the top-left and bottom panels of three possible steady-state syntheses. The synthesis in the top right is the transition from the top-left to the right-left cases. The parameters $(\Gamma,\gamma)$ are respectively taken to be (1.90,0.5), (1.80,1), and (1.69,1.5), from top to bottom and left to right. The color code and the small insert indicate the control law used according to the position of the $M$ point in the Bloch ball. The Bloch limit represents the circle of equation $y^2+z^2=1$. The times 0 and $T$ correspond respectively to the initial and final times of the control field. The different quantities are unitless. The different quantities are unitless.}
\end{figure}

\begin{figure}[htbp]
        \centering\includegraphics[scale=0.5]{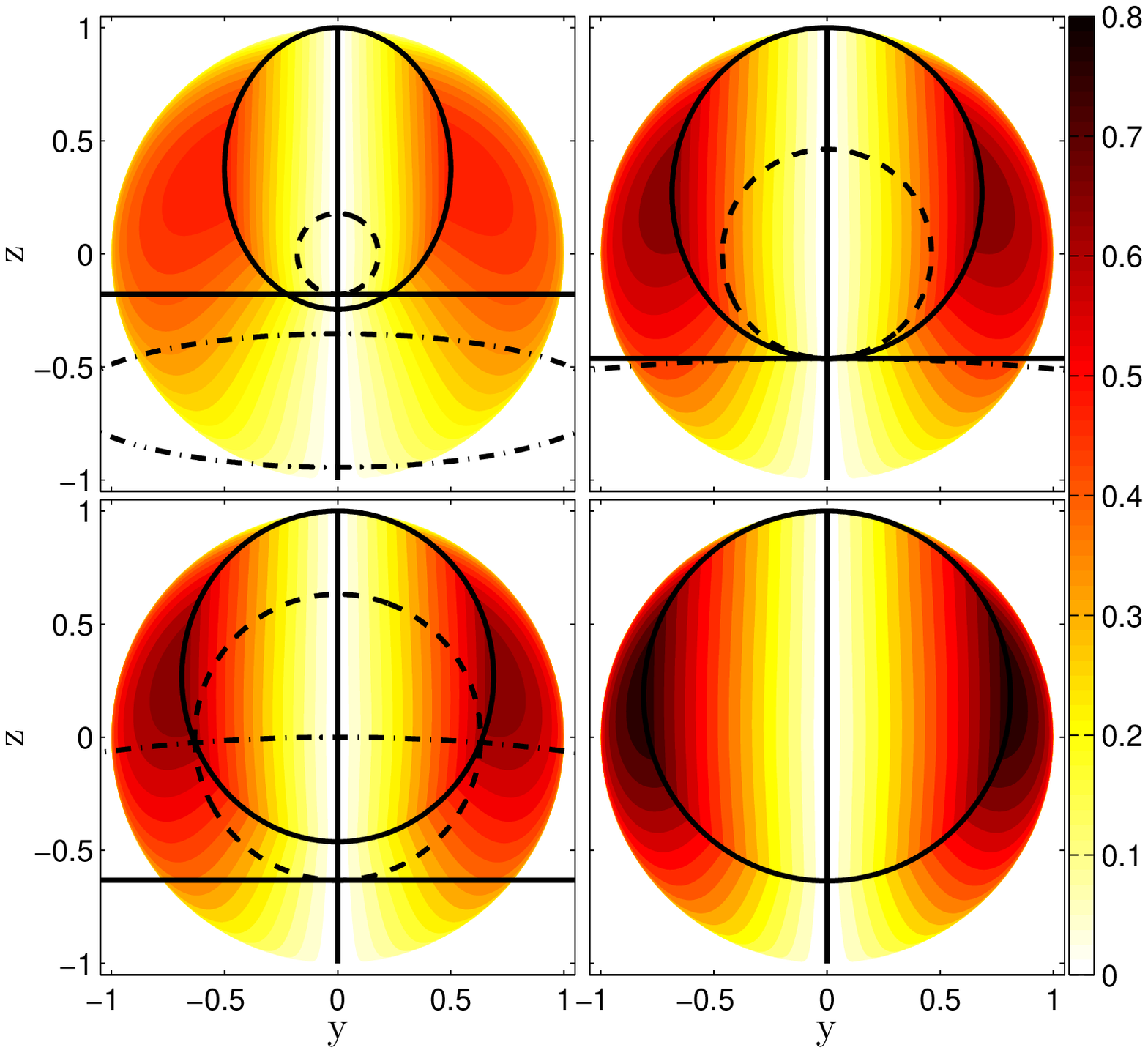}
        \caption{\label{fig5}(Color online) Figure of merit surface $Q(y_m,z_m)$ associated with the steady-state syntheses of Fig.~\ref{fig4}. The solid, dashed and dot-dashed lines represent the boundaries between the different regions of the Bloch ball where the control law is not the same. The definition of these lines is given in Fig.~\ref{fig4}.}
\end{figure}

We have now all the tools in hand to optimize the SNR. This optimization is a non-standard and difficult control problem for which a cost functional $Q$ has to be maximized, together with the determination of the initial $S$ and final $M$ points  of the control period. Only the position of one of the two points needs to be computed since the $S$ and the $M$ points are connected by a free evolution. Following Refs.~\cite{Lapert2014,Lapert2015}, we adopt a brute-force strategy to solve this problem, which consists in replacing this global optimization by a two-step procedure. For any $M$ point of coordinates $(y_m,z_m)$ of the Bloch ball, we first determine as explained in Sec.~\ref{move} the time-optimal path going from $S$ to $M$ and we compute the corresponding value of the figure of merit $Q$. Only five types of control sequences can be optimal. According to the values of the relaxation parameters, only three different optimal syntheses are obtained, as displayed in Fig.~\ref{fig4}. We observe the occurring of the Ernst ellipsoid, which is the set of points where $R_m=R_s$ (the radii of the $M$ and $S$ points), i.e. the control sequence where only a $\delta$- pulse is used to bring the system from $S$ to $M$.
In this unbounded situation, the functional $Q$ can also be determined analytically as a function of the coordinates of $M$~\cite{Lapert2014,Lapert2015}. Mathematically, $Q$ is a non-smooth but continuous surface. This surface is represented in Fig.~\ref{fig5} for the four cases of Fig.~\ref{fig4}.
A gradient method gives the maximum of this surface, which belongs to the Ernst ellipsoid. This point can be characterized by the well-known Ernst angle solution:
\begin{equation}
\cos\theta=\frac{e^{-\gamma}+e^{-\Gamma}}{1+e^{-\Gamma-\gamma}},
\end{equation}
which was first derived in the original paper by Ernst and his co-workers \cite{ernst1966}. The angle $\theta$ denotes here the angle characterizing the $\delta$- pulse of the Ernst sequence.

This first result of optimality on the SNR per unit time paves the way in a near future to a systematic analysis of the
optimization of this parameter in spin systems. This method can be generalized to the case of bounded control amplitudes, the use of crusher gradient pulses, to an inhomogeneous ensemble of spins with magnetic field broadening, or to any other experimental constraint. Other aspects will be discussed in the conclusion of the paper.

\section{Numerical optimal control of the contrast in MRI}\label{sec3}
In this section, a numerical solution of the contrast optimization problem is presented.
Contrast in MRI is a key parameter to visualize and interpret the anatomy and bio-chemical properties of potentially malign tissues.
In clinical practice, contrast is usually empirically handled by tuning few acquisition parameters.
This approach is however strongly radiologist dependent, only provides limited contrast combinations and offers no guarantee about the optimal nature of the created contrast.
The objective of this study is to compute the radio-frequency (RF) magnetic pulse which optimally combines the effect of excitation and relaxation of the spins to produce the maximal signal difference between 2 tissues with different relaxation times.
We recall that signal in MRI is directly linked to the norm of the transverse magnetization vector $\vec{M}$, denoted $\vec{M}_\perp=(M_x,M_y,0)$. Geometric approaches have been proposed to solve this problem for different tissues for spins precessing at the nominal Larmor frequency \cite{Lapert2012,Bonnard2012,Bonnard2013}.
For the application on a realistic MRI system, it is essential to account for magnet imperfections which induce various resonance frequency offsets. In this case, each isochromat is characterized by a magnetization vector $\vec{M}(\omega)=(M_x(\omega),M_y(\omega),M_z(\omega))$ whose dynamics is governed by the following differential system:
\be
\lgr\ba{rl}
\dot{M}_x(\omega) &= -\omega M_y(\omega)-2\pi M_x(\omega)/T_2 + \omega_y M_z(\omega)\\
\dot{M}_y(\omega) &= \omega M_x(\omega)- 2\pi M_y(\omega)/T_2 - \omega_x M_z(\omega)\\
\dot{M}_z(\omega) &= 2\pi (M_0 - M_z)/T_1 - \omega_y M_x(\omega) + \omega_x M_y(\omega),
\ea\rd
\ee
where the offset term $\omega$ belongs to a given interval $[\omega_{min},\omega_{max}]$. From a numerical point of view, this control problem is solved by digitizing the distribution of the isochromats. However, the resulting dimensionality increase (several thousands of spins) makes the geometric solution inapplicable, and requires the use of numerical approaches to solve the problem. We also mention that a similar robustness approach can be applied for $T_1$ and $T_2$ deviations as well as control field inhomogeneities. This section details an example of a contrast experiment, from the optimal control formulation to the \textit{in vivo} acquisition of a rat brain image.
\subsection{Numerical Optimization}
Several numerical schemes have been proposed to solve optimal control problems, including shooting methods \cite{Bonnard2014,Bonnard2014a}, Krotov methods \cite{Vinding2012} and gradient ascent (or descent) based approaches \cite{Khaneja2005, Fouquiere2011}.
This work illustrates a practical application of the gradient ascent pulse engineering (GRAPE) algorithm.
The application of the algorithm requires the temporal discretization of the control field, where each time step represents an optimization variable.
In its simplest version, this algorithm updates each control field time step by minimizing the user-defined cost function at each iteration, while fulfilling the constraints imposed by the Pontryagin Maximum Principle (PMP).
The cost function minimization is performed by applying a step in the direction of the cost function gradient with respect to the control field. The reader is referred to \cite{Khaneja2005,Fouquiere2011} for explicit details on the gradient computation.
Convergence is reached when the gradient and/or the step norms are bellow a user-defined threshold.
Note that unlike the geometric methods, the global nature of the minimum cannot be claimed and global or local minimum is only reached within a certain tolerance.
This tolerance is set in accordance with the experimental requirements so that additional improvements of the control field have a negligible impact on the experimental result.
The control field initialization also impacts the nature of the minimum.
Multiple trials or prior insights on the solution can be used to find the global minimum and to improve the convergence of the algorithm.
\\The cost function ($\mathcal{C}$) that optimizes the contrast must maximize the difference between the transverse magnetization norms of 2 samples $a$ and $b$, at the end of the control time ($t=t_f$). The two species are characterized by different relaxation times $T_1$ and $T_2$. It is arbitrarily chosen here that the signal of the samples $a$ and $b$ are respectively maximized and minimized. Note that other definitions of the contrast could be chosen.
Static field inhomogeneities lead to different resonance frequency offsets which in turn result in different trajectories in the Bloch ball, that must be accounted for in the definition of the cost function.
This can be done by sampling the frequency offset interval in $N$ samples, where each point corresponds to a given trajectory.
The cost function thus enforces all trajectories in the considered interval to have similar behaviors:
\be
\mathcal{C} = \frac{1}{N}\sum_{i=1}^N \left[\Vert \overrightarrow{M_b}_\perp^{(i)}(t_f) \Vert - \Vert \overrightarrow{M_a}_\perp^{(i)}(t_f) \Vert \right]
\ee

\subsection{Example of the rat muscle/brain contrast}
\begin{figure}[]
        \centering\includegraphics[scale=0.2]{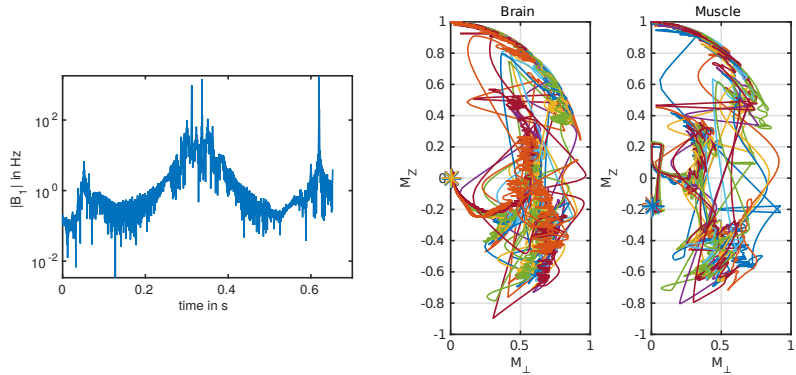}
        \caption{\label{fig6}(Color online) Left: Amplitude of the optimal RF pulse in a log scale graph. Right: Magnetization trajectories of the brain and muscle tissues. Each trajectory represents a specific resonance frequency offset in the interval [-400 Hz, 400 Hz]. A sampling rate of 40 Hz is considered to discretize this interval. Star markers represent the ending point (at time $t=t_f$) of each trajectory.}
\end{figure}

\begin{figure}[]
        \centering\includegraphics[scale=0.3]{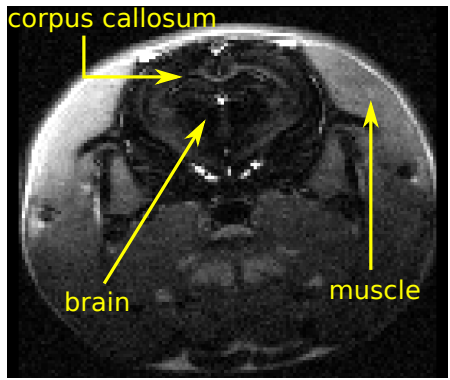}
        \caption{\label{fig7} Transverse section of a rat brain showing the result of the optimal pulse. The aim of the control is to saturate the brain while maximizing the intensity of the surrounding muscle tissues.}
\end{figure}

To illustrate the impact of optimal contrast pulses, a RF pulse is computed to optimize the contrast between the rat brain (minimize) and surrounding muscles (maximize). This example was chosen as a proof-of-concept because it cannot be created with standard contrast strategies, due to the short transverse relaxation time ($T_2$) of the muscle tissues \cite{Bernstein2004}. The following experiment is performed in agreement with the UCBL's ethic committee on animal experimentation, on a 4.7~T Bruker MR system.
Average relaxation times of both structures, as well as the offset inhomogeneity range are estimated by standard MR sequences \cite{Levitt2008}.
Longitudinal ($T_1$) and transverse ($T_2$) relaxation times are respectively estimated to be $[920,\,60]$ ms for the brain and $[1011,\,30]$ ms for the muscle.
The resonance offset range is of the order of 1~kHz. In order to handle slice selectivity, i.e. the location of the image section plane~\cite{Bernstein2004}, standard MRI excitation schemes are used~\cite{Bernstein2004,Reeth2017}.
The optimal pulse is thus used as a contrast preparation pulse, implying that the contrast is prepared on the longitudinal axis ($M_z$).
Only a slight change in the cost function is required to apply this modification~\cite{Reeth2017}.
The prepared magnetization is subsequently flipped into the transverse plane with a slice-selective $\pi$/2-pulse.
We recall that the MRI signal is proportional to the amount of transverse magnetization, i.e. the tissue with the highest $\left|M_z\right|$ at the end of the preparation will produce higher intensity in the MR image.
The magnitude of the resulting optimized pulse is shown in Fig.~\ref{fig6}, together with the magnetization trajectories of the contrasted tissues.
Magnetization trajectories are represented in a plane formed by the normalized transverse ($M_\perp$) and longitudinal ($M_z$) magnetization.
In this figure, different trajectories represent a specific frequency offset.
Notice how all trajectories of a given tissue start from the thermal equilibrium ($\overrightarrow{M_0}$ = (0, 0, 1)) and reach the same final magnetization state despite important trajectory disparities.
This validates the pulse robustness against resonance offset variations.
It can be observed that the brain is saturated at the end of the pulse, i.e. that trajectories reach the center of the Bloch sphere, while a significant amount of longitudinal magnetization is left for the muscle tissues.
This implies that unlike muscle tissues, brain tissues will have a very low contribution to the acquired MRI signal.
The corresponding acquired MR image is shown in Fig.~\ref{fig7}.
As expected, the average tissue intensity within the brain is much lower than the surrounding muscles.
Interestingly, some internal brain structures can be distinguished.
It corresponds to structures whose relaxation times slightly differ from the major brain components which are mostly gray matter.
In particular, the corpus callosum, the largest white matter structure in the brain, can be observed as an hypersignal element.
As white matter has a lower $T_2$ than gray matter \cite{Graaf2006}, this validate the ability of the optimal pulse to enhance short $T_2$ tissues.
This particular contrast opens interesting perspectives in both pre-clinical and clinical neuro-imaging applications since white matter usually appears in hyposignal due to its short $T_2$.
Such a contrast, keeping in mind its optimal nature, could help to improve the visualization of white matter in the human brain, which is a key factor in actual clinical challenges such as the effect of aging or Alzheimer disease~\cite{Black2009}. This approach could also be used to highlight the differences between the structures of the brain.
\section{Conclusion}\label{sec4}
In MRI, there exists an immense potential for improvement as well
as for reduction of the energy deposit and time a patient has to spend in a scanner during an examination. In this work, we have presented two benchmark examples showing the efficiency of this technique. In the ideal case of a homogeneous ensemble of spin systems, we have demonstrated the optimality in the limit of unbounded controls of the Ernst angle solution, derived in the sixties, to enhance the SNR. Numerical optimization techniques were used to illustrate the possibility of maximizing the contrast of the image in MRI by using a particular pulse sequence. This approach was applied with success theoretically and experimentally on the brain of a rat. We emphasize that one of the main advantages of this contrast enhancement is its general character since the optimal control magnetic fields can be computed with standard routines published in the literature and implemented on a MRI scanner without requiring specific materials and process techniques. An important issue is the use of
this method in clinics to optimize the contrast of human body imaging. We anticipate that this technique could be
a complementary tool to contrast agents whose successful application has been an important aspect of the development of MRI in recent years. The joint use of these two methods could therefore limit the concentration of contrast agents needed for the imaging, which could be beneficial to the patient.

These two examples validate the potential of optimal control as an accurate pulse design tool to solve non-trivial practical problems in MRI. This ability can be extended to other contexts in MRI, as shown very recently in \cite{Lefebvre2017} with the introduction of optimal control pulses for magnetization phase control in MRI.
There is a huge interest for accurate phase control in many important applications in MRI, such as diffusion imaging, thermometry and elastography, to improve the signal sensitivity to specific bio-markers.

The use of optimal control for these applications relies on further developments in robust optimal control analytic and numerical tools, accurate modeling of the MR physics and tailored acquisition sequences, that will further improve existing MR methods and approach the system's physical limits.

\section*{Acknowledgements}
  This work is supported by the ANR-DFG research program Explosys (Grant
No. ANR-14-CE35-0013-01; GL203/9-1) and from the Technische Universit\"at
M\"unchen Institute for Advanced Study, funded by the German Excellence
Initiative and the E. U. Seventh Framework Programme under Grant No.
291763. This work was performed within the framework of the LABEX
PRIMES (ANR-11-LABX-0063/ ANR-11-IDEX-0007). M.L. acknowledges support
from the Bayerische Forschungsstiftung.


\end{document}